\documentclass[conference]{IEEEtran}
\IEEEoverridecommandlockouts
\usepackage{etoolbox}
\usepackage{cite}
\usepackage{amsmath,amssymb,amsfonts}
\usepackage{mathrsfs}
\usepackage{algorithm}
\usepackage{algpseudocode}
\usepackage{graphicx}
\usepackage{textcomp}
\usepackage{xcolor}
\usepackage[inkscapelatex=false]{svg}

\usepackage{hyperref}

\def\BibTeX{{\rm B\kern-.05em{\sc i\kern-.025em b}\kern-.08em
    T\kern-.1667em\lower.7ex\hbox{E}\kern-.125emX}}
\begin{document}

\title{Speech Emotion Recognition using Attention-based LSTM-Network with Residual Connection}

\author{\IEEEauthorblockN{Daniil Krasnoproshin\,\IEEEauthorrefmark{1} 
and Maxim Vashkevich\,\IEEEauthorrefmark{2}}
\IEEEauthorblockA{Embedded System Department of\\ Belarusian State University of Informatics and Radioelectronics\\
6, P.~Brovky st., 220013, Minsk, Belarus\\
Email: \IEEEauthorrefmark{2}vashkevich@bsuir.by, \IEEEauthorrefmark{1}daniil.krasnoproshin@gmail.com}}
 
\maketitle

\begin{abstract}
Speech emotion recognition is an important component of modern human–computer interaction systems. However, many state-of-the-art approaches rely on large pretrained models with high computational and memory requirements, limiting their applicability. This paper proposes \textsc{ResLSTM-SA}, a lightweight architecture that integrates residual connections with soft attention within an LSTM-based framework. Evaluated on the RAVDESS dataset under strict speaker-independent partitioning, the proposed model outperforms conventional attention-based LSTM baselines and several previously reported CNN- and hybrid CNN--LSTM architectures in terms of unweighted average recall (UAR). The best-performing variant (\textsc{ResLSTM-SA-h64}) achieves a maximum UAR of 0.6517 with only 46.8\,k trainable parameters, delivering competitive accuracy with three orders of magnitude fewer parameters than large-scale self-supervised alternatives, thereby enabling efficient deployment on edge devices and real-time voice assistants.
The source code is available at \url{https://github.com/Mak-Sim/ResLSTM-SER}.
\end{abstract}


\section{Introduction}
Speech emotion recognition (SER) has emerged as a rapidly advancing research field due to its critical role in diverse applications including human-computer interaction, healthcare, customer service in call centers, and affective e-learning systems~\cite{George-2024, Feng-2023}.
Deep learning models have become a mainstream approach for SER system development~\cite{Luna-2021,Dissanayake-2020,Luna-2022}. While research efforts have increasingly shifted toward multimodal emotion recognition systems~\cite{Luna-2021, Luna-2022, Lee-2025} that fuse audio and visual modalities, numerous real-world applications -- such as voice assistants and customer service platforms -- are inherently constrained to the audio modality.
Pretrained models such as PANNs (large-scale pretrained audio neural networks)~\cite{Kong-2020} and Wav2Vec~\cite{Baevski-2020,Luna-2022} achieve high accuracy for SER task; however, their considerable computational cost limits their applicability in real-world settings.

This work aims to develop an efficient SER system that achieves a trade-off between computational complexity and recognition accuracy. We propose a lightweight LSTM-based architecture supplemented with a soft attention mechanism and residual connections to enhance temporal modeling while maintaining low computational load. The remainder of this paper is organized as follows. Section 2 describes the acoustic feature extraction pipeline. Section 3 details the proposed LSTM-based architecture and its variants. Section 4 presents the experimental setup and results. Finally, Section 5 concludes the paper and discusses directions for future work.
\begin{figure*}[h]
\centering
\includesvg[width=0.92\linewidth]{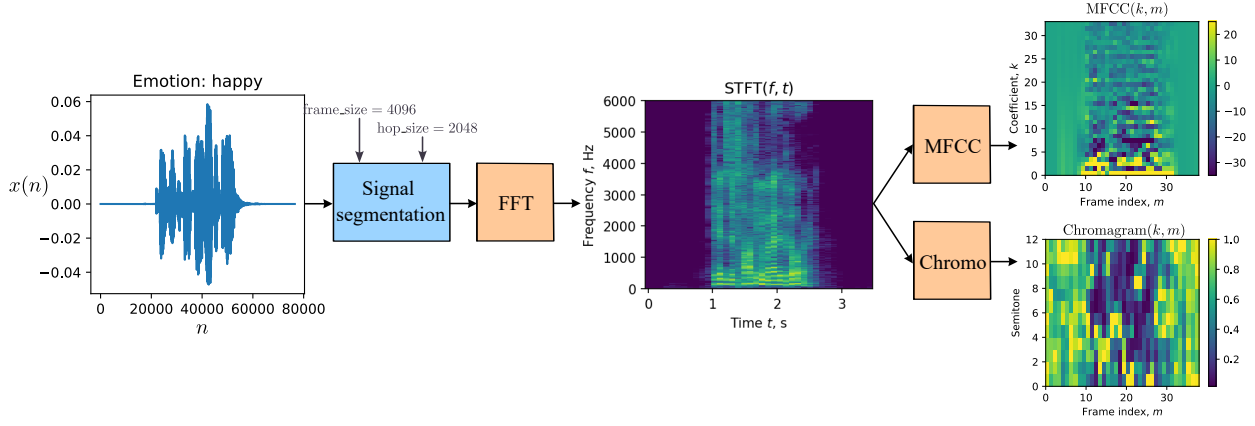}
\caption{Feature extraction process}
\label{fig:feature_extraction}
\end{figure*}

\section{Feature extraction}
Raw speech signals encode rich information about expressed emotions, however, to be processed by deep learning models, this information must be transformed into a low-dimensional representation. In this work, we parameterize the speech signal using two popular techniques: mel-frequency cepstral coefficients (MFCCs) and chromagrams.

Mel-frequency cepstral coefficients (MFCCs) are derived from psychoacoustic principles that approximate the human auditory system's nonlinear frequency resolution. This representation converts a raw audio signal into a sequence of compact, fixed-dimensional feature vectors suitable for machine learning. The computation proceeds in several stages. First, the signal is segmented into overlapping frames. For each frame, the short-time Fourier transform (STFT) is applied to obtain a time--frequency representation:
\begin{equation}
X(m,k) = \sum_{n=0}^{N-1} x(n + m h_{\mathrm{size}})\, w(n)\, e^{-j2\pi kn/N},
\end{equation}
where $x(\cdot)$ denotes the input signal, $w(n)$ is a window function, $N$ is the frame length, $h_{\mathrm{size}}$ is the hop size, $m$ indexes the frame, and $k$ indexes the frequency bin. Subsequently, the magnitude spectrum is mapped onto the mel scale using a triangular filterbank, followed by logarithmic compression and a discrete cosine transform (DCT) to yield the final mel-frequency cepstral coefficients.

To enrich the speech representation with prosodic cues critical for SER, we incorporate chroma features~\cite{ellis2007chroma}. Chroma features have been employed in several recent studies for speech emotion recognition~\cite{Tawfeeq-25,Das-24}. Supplementing conventional acoustic representations such as MFCCs and Mel-spectrograms with chroma features has been shown to yield significant performance improvements.

A chromagram is a time-frequency representation that projects the spectral content onto the 12 pitch classes of the chromatic scale, thereby capturing harmonic structure while remaining invariant to octave shifts. Chroma features are derived from the magnitude spectrogram $|X(m,k)|$ by mapping each frequency bin to one of the 12 chroma bins according to its pitch class. The resulting 12-dimensional chromagram provides a compact representation that encodes pitch-class information relevant to vocal intonation and emotional prosody.

In this study, we extracted 34-dimensional MFCC vector and 12-dimensional chroma vector per frame and concatenated them to obtain a 46-dimensional joint representation. The feature extraction pipeline is illustrated in Fig.~\ref{fig:feature_extraction}.

As a result, the speech signal is transformed into a temporal sequence of feature vectors:
\begin{equation}
\mathbf{X} = \left[ \mathbf{x}_1, \mathbf{x}_2, \dots, \mathbf{x}_T \right]^\top \in \mathbb{R}^{T \times d},
\end{equation}
where $T$ denotes the sequence length (number of frames) and $d = 46$ is the feature dimensionality. This matrix representation $\mathbf{X}$ serves as the input to a recurrent neural network.

\section{LSTM-based networks for SER}
This work is based on the LSTM-based architecture with a soft attention mechanism proposed in~\cite{Mirsamadi-2017}. The LSTM cell processes sequential inputs through gated interactions governed by the following equations:
\begin{equation*}
\begin{aligned}
\mathbf{i}_t &= \sigma\!\left( \mathbf{W}_{ii} \mathbf{x}_t + \mathbf{b}_{ii} + \mathbf{W}_{hi} \mathbf{h}_{t-1} + \mathbf{b}_{hi} \right) \\
\mathbf{f}_t &= \sigma\!\left( \mathbf{W}_{if} \mathbf{x}_t + \mathbf{b}_{if} + \mathbf{W}_{hf} \mathbf{h}_{t-1} + \mathbf{b}_{hf} \right) \\
\mathbf{g}_t &= \tanh\!\left( \mathbf{W}_{ig} \mathbf{x}_t + \mathbf{b}_{ig} + \mathbf{W}_{hg} \mathbf{h}_{t-1} + \mathbf{b}_{hg} \right) \\
\mathbf{o}_t &= \sigma\!\left( \mathbf{W}_{io} \mathbf{x}_t + \mathbf{b}_{io} + \mathbf{W}_{ho} \mathbf{h}_{t-1} + \mathbf{b}_{ho} \right) \\
\mathbf{c}_t &= \mathbf{f}_t \odot \mathbf{c}_{t-1} + \mathbf{i}_t \odot \mathbf{g}_t \\
\mathbf{h}_t &= \mathbf{o}_t \odot \tanh(\mathbf{c}_t) ,
\end{aligned}
\end{equation*}
where $\mathbf{x}_t \in \mathbb{R}^d$ is the input feature vector at time step $t$, $\mathbf{h}_t$ and $\mathbf{c}_t$ denote the hidden state and cell state, respectively, $\mathbf{i}_t$, $\mathbf{f}_t$, $\mathbf{g}_t$, and $\mathbf{o}_t$ represent the input, forget, cell candidate, and output gates, $\sigma(\cdot)$ is the sigmoid activation function, $\odot$ denotes the element-wise product, and $\mathbf{W}_{\ast\ast}$ and $\mathbf{b}_{\ast\ast}$ are trainable weight matrices and bias vectors.

In~\cite{Mirsamadi-2017} to make utterance-level emotion classification the sequence of hidden states $\mathbf{h}_t$ is aggregated using attention mechanism and fed into fully-connected layer with $\mathrm{softmax}$ activation function.
The context vector is computed as a weighted sum of the hidden states:
\begin{equation}
\mathbf{h}_{\mathrm{context}} = \sum_{t=0}^{T-1} \alpha_{t} \mathbf{h}_{t},
\end{equation}
where $\alpha_t$ denotes the attention weight for time step $t$, representing the relative importance of the corresponding hidden state $\mathbf{h}_{t}$ in the utterance-level representation.

A soft attention mechanism was used to obtain attention coefficients:
\begin{equation}
\alpha_{t} = \mathrm{softmax}(e_{t}) = \frac{\exp(e_{t})}{\sum_{t=0}^{T-1} \exp(e_{t})},
\end{equation}
where $e_{t} = \mathbf{u}^{T} \mathbf{h}_{t}$ is the attention score, and $\mathbf{u}$ is the attention vector (trainable parameter).
\begin{figure}[!h]
\centering
\includesvg[width=0.75\linewidth]{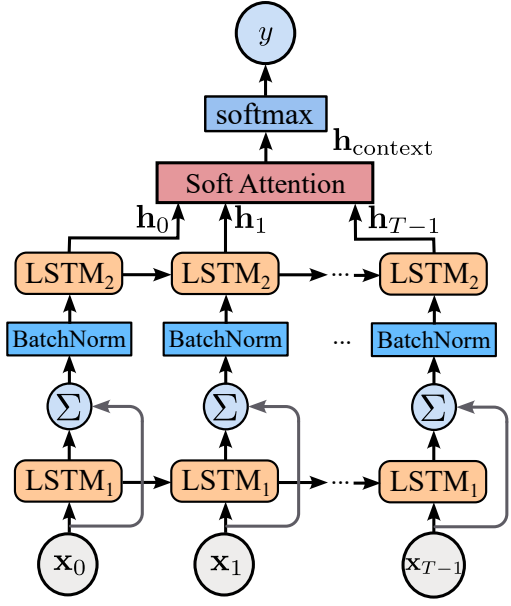}
\caption{The scheme of the ResLSTM-SA model}
\label{fig:ResLSTM}
\end{figure}

We denote the described architecture~\cite{Mirsamadi-2017} with a hidden state dimensionality of $Y$ as \textsc{LSTM-SA-h}$Y$, where ``SA'' denotes soft attention.

In this paper, we propose a modification to the LSTM-SA architecture by incorporating an additional LSTM layer with skip connection. This layer enriches the temporal representation of the input features prior to processing by the main attention-based LSTM, thereby enhancing the model's capacity to capture long-range dependencies. The scheme of the proposed architecture is shown in Fig.~\ref{fig:ResLSTM}.
We denote the proposed architecture as \textsc{ResLSTM-SA}, where ``Res'' refers to residual connections. To specify models with a hidden state dimensionality of $Y$, we adopt the notation \textsc{ResLSTM-SA-h}$Y$ (e.g., \textsc{ResLSTM-SA-h}128 for $Y=128$).

\begin{figure}[!htb]
  \centering
  \includegraphics[width=.7\linewidth]{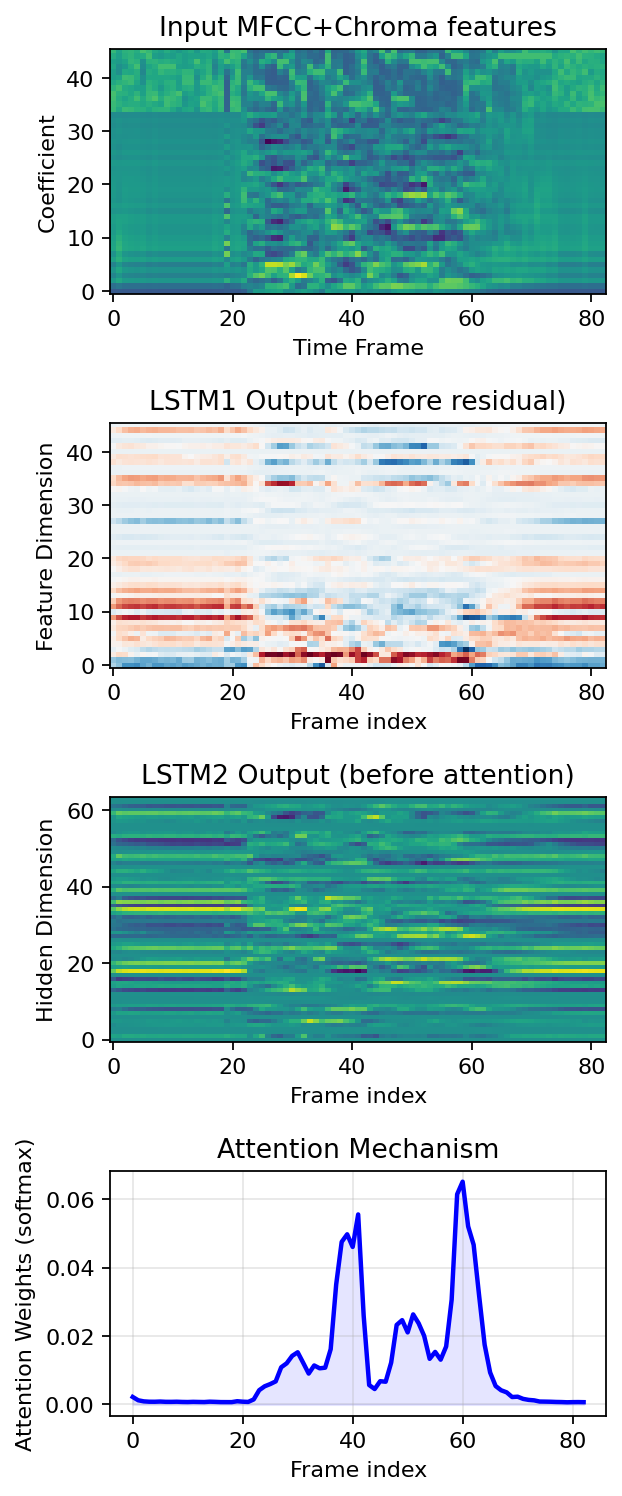}
  \caption{Visualization of signal flow in \textsc{ResLSTM-SA-h64}}
  \label{fig:ResLSTM_hidden_state}
\end{figure}

A distinctive characteristic of the proposed architecture is that the hidden state dimensionality of the first LSTM layer matches the input feature dimension ($d=46$). This design enables residual connections that allow the model to enrich the input sequence with contextual information while preserving the original feature structure.
Fig.~\ref{fig:ResLSTM_hidden_state} visualizes the input feature sequence and intermediate activations of the trained \textsc{ResLSTM-SA-h64} model.

The figure illustrates how $\mathrm{LSTM}_1$ generates contextually enriched representations that are added to the original input via residual connections. Furthermore, the attention mechanism concentrates model focus on salient temporal regions around frames 40 and 60, corresponding to emotionally expressive segments of the utterance.

By maintaining input-matched dimensionality in the first LSTM layer, residual connections enable additive fusion of raw features with their contextually enriched counterparts. This preserves initial information while allowing the attention-based LSTM to focus exclusively on higher-order emotional pattern extraction.

\section{Experiments}

\subsection{Dataset}
The study used the emotional speech audio subset of the Ryerson Audio-Visual Database of Emotional Speech and Song (RAVDESS)~\cite{Livingstone-2018}. This subset comprises 1,440 high-quality audio files (16-bit, 48 kHz), featuring 24 professional actors (12 male, 12 female), reciting 2 statements ``Kids are talking by the door'' and ``Dogs are sitting by the door'' with 8 different emotions. The expressed emotions include neutral, calm, happy, sad, angry, fearful, surprised, and disgusted. To capture intensity variance, all emotional expressions (excluding neutral) were produced at two levels of intensity (normal and strong). Each utterance was produced twice.

\subsection{Experimental setup}
The core objective of this study was to conduct a comparative evaluation of LSTM-SA and ResLSTM-SA architectures for SER. The experimental design systematically assessed the impact of two architectural factors: (i) the inclusion of residual connections and (ii) model capacity as controlled by hidden state dimensionality $h \in \{32, 64, 128\}$. All models were trained using the Adam optimizer with a cosine annealing learning rate scheduler (with period $T_0$ epochs), which adaptively modulates the learning rate to escape local minima and improve convergence stability.

Model weights were initialized using Xavier normal initialization to preserve activation variance during forward and backward propagation. LSTM gate biases were set to zero except for the forget gate, whose bias was initialized to one to ensure the initial storage of information in the memory cell.

The categorical cross-entropy loss was employed, the standard objective for multi-class classification that yields well-behaved gradients when combined with softmax activation:
\begin{equation}
\mathcal{L}_{\mathrm{CE}} = -\frac{1}{N} \sum_{n=1}^{N} \sum_{c=1}^{C} y_{n,c} \, \log \bigl( \hat{y}_{n,c} \bigr),
\end{equation}
where $N$ denotes the batch size, $C$ the number of emotion classes, $y_{n,c} \in \{0,1\}$ the ground-truth label (one-hot encoded), and $\hat{y}_{n,c} \in (0,1]$ the predicted class probability from the softmax layer. The negative sign ensures minimization of the Kullback--Leibler divergence between true and predicted distributions.

\subsection{Hyperparameter optimization and performance estimation}
A Bayesian hyperparameter optimization framework~\cite{Akiba-2019} called Optuna was employed to ensure a fair comparison by identifying the optimal training configuration for each model variant. This approach dynamically generates the search space during optimization. Optuna uses the Tree-Structured Parzen Estimator (TPE) as its hyperparameter sampling algorithm. TPE is a Bayesian optimization method that models the probability densities of high- and low-performing hyperparameter configurations separately. By comparing these distributions, TPE intelligently directs the search to promising regions of the hyperparameter space, significantly accelerating convergence compared to random or grid search methods.

Hyperparameter optimization involved five main parameters: learning rate (log-sampled in the range
$\eta  = [3 \cdot 10^{-5}, 2 \cdot 10^{-4}]$), weight decay (log-sampled in the range
$\lambda =[2 \cdot 10^{-5}, 2 \cdot 10^{-2}]$), dropout in fully connected layers (uniform sampling in the range ${p}_{drop} = [0.1, 0.5]$), the number of annealing cycles in the cosine annealing scheduler $T_0 \in \{1, 2, 3, 5, 10\}$ and  $\mathrm{batch\_size} \in \{8, 16, 32, 64\}$. Each model was trained for 100 epochs, as accuracy on the validation set showed virtually no improvement after that.

To obtain a robust and unbiased estimate of model performance, we performed 10 independent training runs using the Optuna-identified optimal hyperparameter configuration, each initialized with a distinct random seed. This protocol mitigates two critical sources of estimation bias: (i) variance arising from stochastic weight initialization, and (ii) potential overfitting to the random seed employed during hyperparameter optimization. The final performance metrics report the mean and standard deviation across these 10 trials, providing a statistically grounded assessment of the architecture's true generalization capability.

Unweighted Average Recall (UAR) was adopted as the evaluation metric. UAR computes the arithmetic mean of per-class recalls, giving equal weight to each emotion category regardless of its prevalence in the dataset:
\begin{equation}
\mathrm{UAR} = \frac{1}{C} \sum_{c=1}^{C} \frac{A_{c,c}}{\sum_{i=1}^{C} A_{c,i}},
\end{equation}
where $A \in \mathbb{N}^{C \times C}$ is the confusion matrix with entry $A_{c,i}$ representing the number of samples from class $c$ predicted as class $i$. The term $A_{c,c} / \sum_{i=1}^{C} A_{c,i}$ corresponds to the recall (true positive rate) of class $c$. 
Model performance was evaluated using 5-fold cross-validation with a speaker-independent partitioning scheme proposed in~\cite{Luna-2021}. This protocol ensures that speakers appearing in the training folds are excluded from the corresponding validation fold, thereby preventing speaker identity leakage and enabling reproducible comparison with prior RAVDESS-based studies.

\subsection{Results}
Table~\ref{tab1:results} summarizes the performance of LSTM-based architectures in terms of UAR. The results demonstrate that incorporating residual connections via the initial LSTM layer in ResLSTM-SA consistently improves recognition accuracy compared to the baseline LSTM-SA architecture, with gains of up to 7.7 percentage points observed across hidden state dimensionalities.
\begin{table}[!h]
\setlength\tabcolsep{3pt}
\caption{Results of experiments with LSTM-based networks with soft attention}
\begin{center}
\begin{tabular}{| l | c | c | c |}
\hline
 \bf Model & \bf \# Params & \bf UAR  &\bf UAR (max) \\ 
\hline
LSTM-SA-h32    & 10.6\;k  & 0.5352 $\pm$ 0.0123 & 0.5547 \\
\hline
LSTM-SA-h64    & 28.3\;k  & 0.5751 $\pm$ 0.0108 & 0.5996 \\
\hline
LSTM-SA-h128   & 91.6\;k  & 0.5895 $\pm$ 0.0076 & 0.6022 \\
\hline
ResLSTM-SA-h32 & 28.0\;k  & 0.6130 $\pm$ 0.0111 & 0.6315 \\
\hline
ResLSTM-SA-h64 & 46.8\;k  & 0.6232 $\pm$ 0.0119 & \textbf{0.6517} \\
\hline
ResLSTM-SA-h128& 108.9\;k & 0.6107 $\pm$ 0.0134 & 0.6348 \\
\hline
\end{tabular}
\label{tab1:results}
\end{center}
\end{table}

For baseline LSTM-SA models, increasing the hidden dimensionality leads to a monotonic improvement in mean UAR, from 0.5352 for $h=32$ to 0.5895 for $h=128$. However, this improvement comes at the cost of a substantial increase in model complexity.

The proposed ResLSTM-SA architectures achieve higher UAR scores while maintaining comparable or even lower parameter counts. For instance, ResLSTM-SA-h32, with only 28.0\;k parameters, outperforms the much larger LSTM-SA-h128 model (91.6\;k parameters), achieving a mean UAR of 0.6130 compared to 0.5895. This corresponds to a relative improvement of approximately 4.0 percentage points in UAR with nearly three times fewer parameters.
The best overall performance is obtained with ResLSTM-SA-h64, which achieves a mean UAR of 0.6232 and a maximum UAR of 0.6517, the highest among all evaluated models. Notably, this configuration achieves an effective balance between representational capacity and regularization, as further increasing the hidden size to $h=128$ does not yield additional gains and instead leads to a slight degradation in mean UAR, likely due to overfitting.

\begin{figure}
\centering
\includesvg[width=0.95\linewidth]{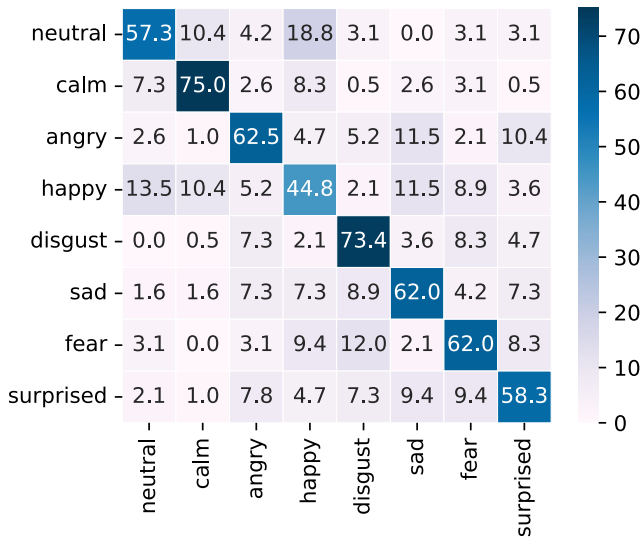}
\caption{Confusion matrix for ResLSTM-SA-h64 model}
\label{fig:ConfMatrix}
\end{figure}

The confusion matrix for the \textsc{ResLSTM-SA-h64} model is shown in Fig.~\ref{fig:ConfMatrix}. Analysis reveals that \textit{happy} exhibits the lowest class-wise recall (44.8\%), primarily due to frequent misclassification as \textit{neutral}. Conversely, 18.8\% of \textit{neutral} utterances are erroneously classified as \textit{happy}, indicating a systematic confusability between these affective states

\begin{figure*}
\centering
\includesvg[width=0.95\linewidth]{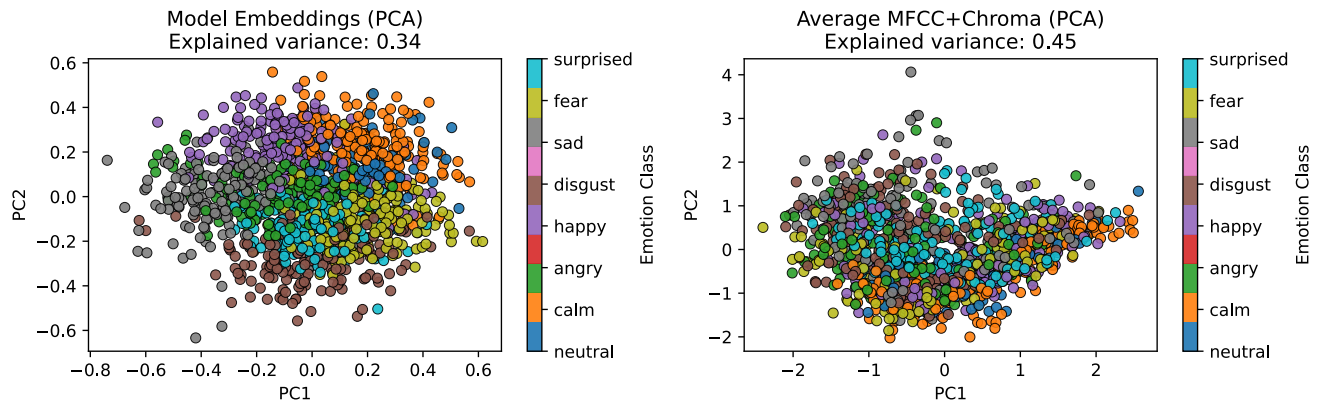}
\caption{PCA visualization of ResLSTM-SA-h64 model embeddings and averaged MFCC+Chroma features}
\label{fig:pca_embeddings}
\end{figure*}

Fig.~\ref{fig:pca_embeddings} visualizes PCA projections of utterance-level representations: (a) embeddings extracted from the \textsc{ResLSTM-SA-h64} model and (b) averaged MFCC+Chroma features. The raw acoustic features exhibit considerable overlap between emotion classes and elevated intra-class dispersion. In contrast, the model-derived embeddings form compact, well-separated clusters with reduced intra-class variance and enhanced inter-class margins. This demonstrates the model's capacity to integrate temporal context and encode prosodic cues into a discriminative latent space that effectively disentangles emotional categories.

Table~\ref{tab:compare} compares the proposed model with published state-of-the-art methods on the RAVDESS dataset. Classical CNN-based and hybrid CNN-LSTM models achieve UAR values in the range of 0.56–0.62, with performance generally increasing alongside model complexity. Notably, the ResLSTM-SA-h64 model surpasses all non–self-supervised approaches, achieving a UAR of 0.6517 while using only 0.05 M parameters, which is three orders of magnitude fewer than fine-tuned CNN and transformer-based models.
\begin{table}[!h]
\setlength\tabcolsep{3pt}
\caption{Comparison results with the state-of-the-art methods
on the RAVDESS dataset}
\begin{center}
\begin{tabular}{|l|c|c|}
\hline
Model & \#Params & UAR \\
\hline
AlexNet embeddings + SVM~\cite{Luna-2021}   & 61.0\;M  & 0.4580 \\
CNN+LSTM~\cite{Dissanayake-2020}            & -        & 0.5671 \\
GResNet+S~\cite{Zeng-2019}                  & -        & 0.5970 \\
Fine-tuned AlexNet~\cite{Luna-2021}         & 61.0\;M  & 0.6167 \\
ResLSTM-SA-h64~[proposed]                   & 0.05\;M  & 0.6517 \\
Fine-tuned CNN14~\cite{Luna-2021}           & 81.0\;M  & 0.7658 \\
Fine-tuned xlsr-wav2vec 2.0~\cite{Luna-2022} & 317.0\;M & 0.8182 \\
wav2vec 2.0 with data augmentation~\cite{Ibrahim-2024} & 317.0\;M & 0.8229 \\
\hline
\end{tabular}
\label{tab:compare}
\end{center}
\end{table}

While large-scale self-supervised models such as PANNs (CNN14)~\cite{Kong-2020} and wav2vec~2.0~\cite{Baevski-2020} achieve higher absolute UAR scores, they require extensive pretraining on massive external corpora and contain between 81\,M and 317\,M parameters. In contrast, the proposed \textsc{ResLSTM-SA} operates as a lightweight, end-to-end trainable architecture with fewer than 50\,k parameters, offering a favorable trade-off between recognition accuracy, inference latency, and model footprint. This efficiency profile is particularly advantageous for edge devices, real-time voice assistants, and scenarios with limited labeled training data -- where computational overhead and memory footprint impose strict operational constraints.

Overall, these results demonstrate that the proposed ResLSTM-SA architecture achieves state-of-the-art performance among compact, task-trained models on RAVDESS. In addition, it exhibits superior parameter efficiency, outperforming larger CNN- and LSTM-based baselines. Moreover, the proposed model architecture provides a competitive alternative to large self-supervised models when model size, training cost, or deployment constraints are primary considerations.

\section{Conclusion}
This paper introduced \textsc{ResLSTM-SA}, a lightweight speech emotion recognition architecture that integrates residual connections with soft attention to achieve an effective trade-off between recognition accuracy and computational efficiency. Evaluated on the RAVDESS dataset under strict speaker-independent partitioning~\cite{Luna-2021}, the proposed architecture consistently outperformed conventional attention-based LSTM baselines across all model capacities ($h \in \{32,64,128\}$). 
Notably, \textsc{ResLSTM-SA-h64} attained a mean UAR of $0.6232 \pm 0.0119$ (max: 0.6517) with 46.8\,k parameters—surpassing the LSTM-SA-h128 baseline by 3.4 percentage points while using approximately half the number of parameters.
Despite operating with three orders of magnitude fewer parameters than large-scale self-supervised models (e.g., wav2vec~2.0 with 317\,M parameters), \textsc{ResLSTM-SA} achieves competitive performance suitable for edge devices and real-time voice assistants.
Future work will extend this architecture to cross-corpus evaluation on challenging benchmarks and explore fusion with lightweight prosodic descriptors to further enhance robustness under acoustic variability.

\bibliographystyle{IEEEtran}
\bibliography{./IEEEabrv,./ser_ref}


\vspace{12pt}

\end{document}